\documentclass[aps,prd,amssymb,nofootinbib,twocolumn,epsf,floatfix]{revtex4}
\usepackage[usenames]{color}
\usepackage[normalem]{ulem} 
\usepackage{graphicx} 
\usepackage{subfigure} 
\usepackage{amssymb}  
\usepackage{amsmath}
\usepackage{mathrsfs} 
\usepackage{epstopdf}

\makeatletter
\renewcommand{\p@subfigure}{\thefigure}
\makeatother

\begin{document}
 

\title{Parametric Representations of Neutron-Star Equations of State
  \\With Phase Transitions}
 
\author{Lee Lindblom}
\affiliation{Department of Physics, University of California at San
  Diego}

\date{\today}
 
\begin{abstract}
  This paper explores the use of low-dimensional parametric
  representations of neutron-star equations of state that include
  discontinuities caused by phase transitions.  The accuracies of
  optimal piecewise-analytic and spectral representations are
  evaluated for equations of state having first- or second-order phase
  transitions with a wide range of discontinuity sizes.  These results
  suggest that the piecewise-analytic representations of these
  non-smooth equations of state are convergent, while the spectral
  representations are not.  Nevertheless, the lower-order ($2\leq
  N_\mathrm{parms}\leq 7$) spectral representations are found to be
  more accurate than the piecewise-analytic representations with the
  same number of parameters.
\end{abstract} 
 
\maketitle 

\section{Introduction}
\label{s:Introduction}

The equation of state of the material in the cores of neutron stars is
not well known at this time.  The density of this material far exceeds
the limits of current laboratory experiments, and there is at present
no universally accepted theoretical model of this material.
Astrophysical observations of neutron stars can in principle be used
to determine the neutron-star equation of state~\cite{Lindblom1992}.
However, the quality and quantity of those observations are presently
quite limited.

Parametric representations of the neutron-star equation of state have
been introduced as a way to analyze the results of the relevant
astrophysical observations.  The parameters in these equations of
state are adjusted to provide best-fit models of the observational
data, thus producing approximate representations of the physical
equation of state.  Since the quality and quantity of the relevant
astrophysical observations are still quite limited, parametric
representations that provide good accuracy using only a small number
of parameters are needed.

Two types of parametric representations of the equation of state have
been introduced for this purpose.  Piecewise-analytic representations,
first introduced in Ref.~\cite{Read:2008iy}, divide the range of
densities into discrete ranges with parameter-dependent analytic
expressions representing the equation of state within each range.
Another type of parametric representation, first introduced in
Ref.~\cite{Lindblom2010}, is constructed from a generating function
expressed as a linear combination of fixed basis functions, e.g.
polynomial or trigonometric functions.  The parameters in these
``spectral'' representations are the coefficients that multiply the
basis functions in the sum that determines the generating function for
the equation of state.

The accuracies of both the piecewise-analytic and the spectral
representations have been evaluated using a diverse collection of
theoretical neutron-star equation of state models~\cite{Read:2008iy,
  Lindblom2010, Lindblom2018, Lindblom2022}.  Those tests showed that
both types of representation are convergent in the sense that their
accuracies increased as the number of parameters in the representation
increased.  Those tests also showed that reasonably good accuracies
(at the few percent level) could be achieved with representations
having a fairly small number of parameters.  Consequently both the
piecewise-analytic and the spectral representations have been widely
used to analyze the presently available observational data, with
Refs.~\cite{Read:2008iy} and \cite{Lindblom2010} having received
hundreds of citations in the literature.

Previous tests of the accuracy of these parametric representations
used a collection of mostly discontinuity-free theoretical equation of
state models.  The physical neutron-star equation of state may (or may
not) include discontinuities caused by phase transitions.  The purpose
of this paper is to systematically evaluate the accuracy of the
piecewise-analytic and the spectral representations when used to
represent non-smooth neutron-star equations of state with phase
transitions.  A sequence of exemplar equations of state are
constructed in Appendix~\ref{s:ExemplarEOS} for this study with phase
transitions having a range of sizes. Those exemplar equations of state
are then used to test the accuracy of both the piecewise-analytic and
the spectral representations.

The methods used in this study to construct optimal parametric
equation of state fits are described in Sec.~\ref{s:OptimalFits}.
These methods are then used to construct optimal fits to each exemplar
equation of state using both the piecewise-analytic and the spectral
representations.  The accuracies of the resulting optimal fits are
evaluated using the $L_2$ norm of the difference between the exemplar
equation of state and its parametric representation.  These results
illustrate how the accuracies of the parametric representations depend
on the type of representation (piecewise analytic or spectral), the
size and type (first- or second-order) of the phase-transition
discontinuities, and the orders of the parametric representations.

Section~\ref{s:Discussion} discusses the implications of the results
found here.  If and when more accuracy is needed to model future
improvements in the quality and quantity of observational data, a
split domain method for constructing more accurate representations of
non-smooth equations of state with phase transitions is proposed.

\section{Optimal Parametric Fits}
\label{s:OptimalFits}

\begin{figure}[!b]
  \centerline{
    \includegraphics[width=0.4\textwidth]{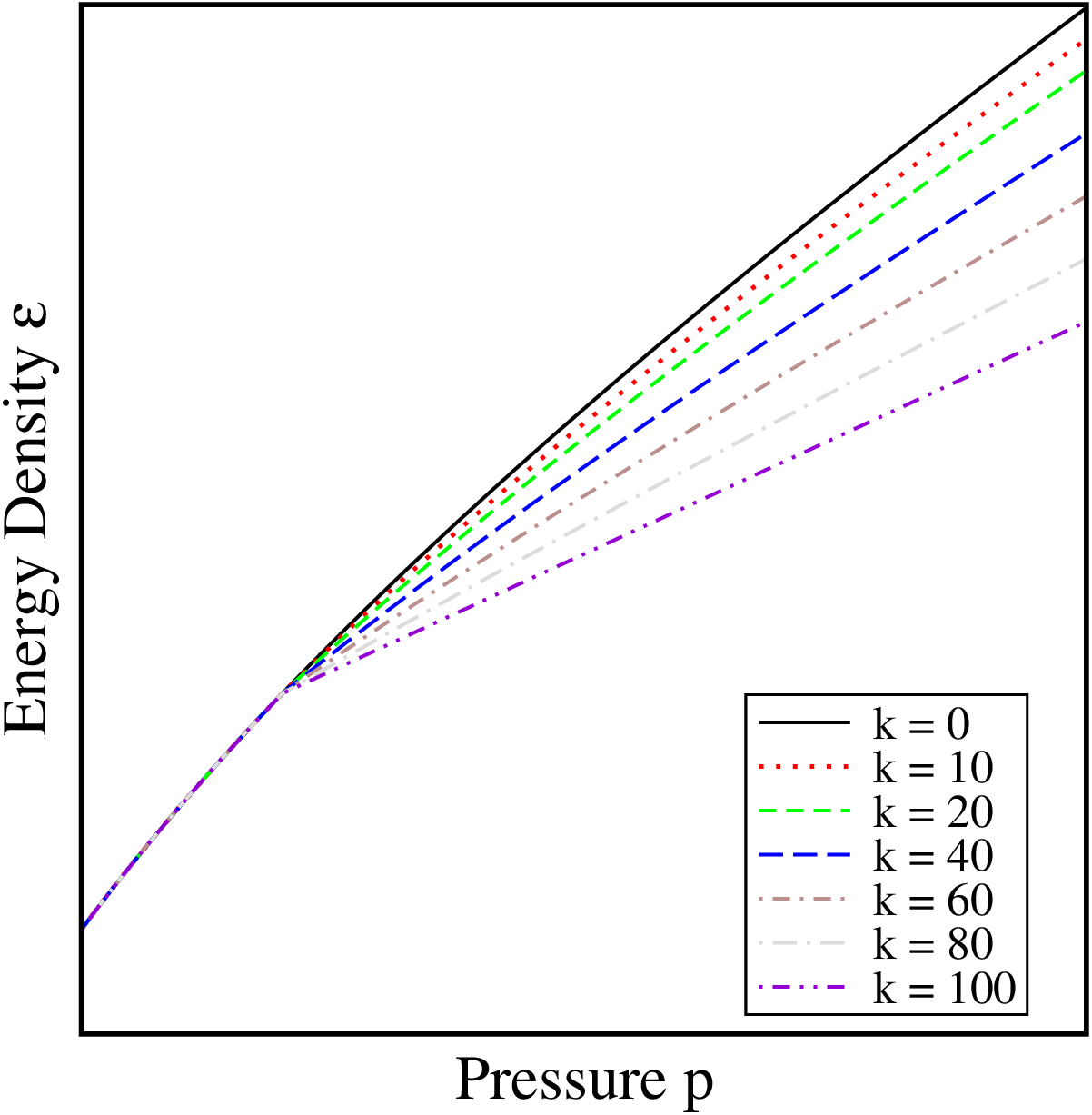}}
  \caption{\label{f:SecondOrderExamples}Several exemplar equations of
    state with second-order phase transitions are illustrated in a
    neighborhood of the phase transition point.  The curves shown here
    include several equations of state with density derivative offsets
    in the range $0\leq k \leq 100$. The $k=0$ curve represents the
    original GM1L equation of state, and the $k=100$ curve has the
    largest physically possible fluid-velocity discontinuity, with the
    fluid velocity equal to the speed of light on the high density
    portion of this curve.}
\end{figure}

This section describes the method used in this study to test the
accuracy of optimal piecewise-analytic and spectral representations of
neutron-star equations of state with phase transitions.  The exemplar
equations of state with phase transitions used to perform these tests
were constructed from the relatively smooth GM1L equation of state,
which is based on a mean-field representation of the interactions
between nucleons~\cite{Typel2010}.  Discontinuities representing
first- or second-order phase transitions were inserted into a
tabulated representation of GM1L at a point several times nuclear
density where the energy density has the value $\epsilon_{\,T}=8\times
10^{14}$ g/cm${}^3$.  These discontinuities were inserted with a range
of sizes specified by a parameter $k$, which determines the size of
the discontinuity as a fraction of the maximum physically relevant
discontinuity (see Appendix~\ref{s:ExemplarEOS}).  The family of
exemplar equations of state used in this study range from the original
smooth GM1L equation of state with $k=0$ to equations of state with
the maximum discontinuity of each type with $k=100$.  Details of the
construction of these exemplar equations of state are given in
Appendix~\ref{s:ExemplarEOS}.  Figures~\ref{f:FirstOrderExamples} and
\ref{f:SecondOrderExamples} illustrate members of these exemplar
equation of state families, with first- and second-order phase
transitions respectively, in the neighborhood of the phase transition
point.
\begin{figure}[!b]
  \centerline{
    \includegraphics[width=0.4\textwidth]{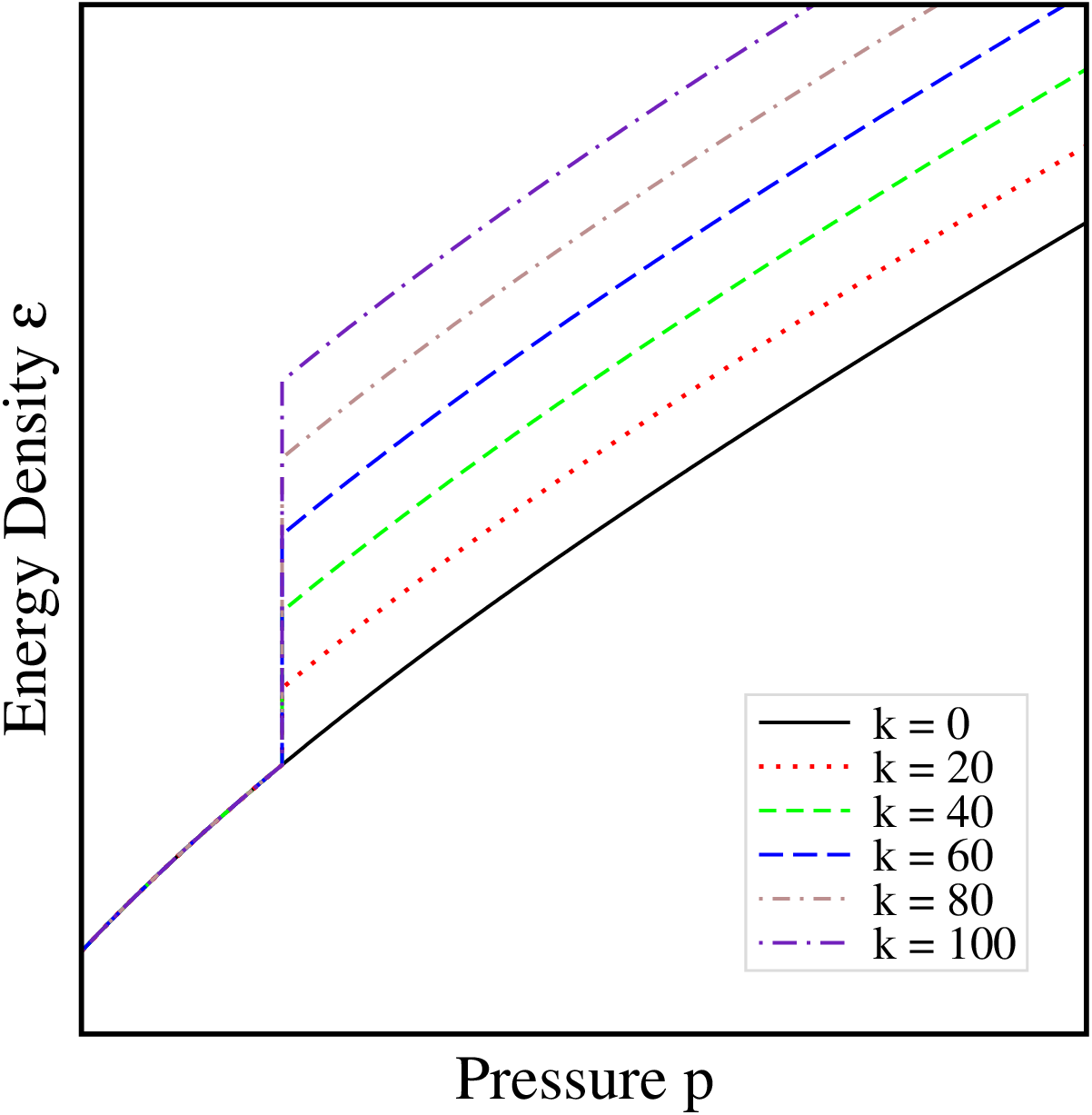}}
  \caption{\label{f:FirstOrderExamples}Several exemplar equations of
    state with first-order phase transitions are illustrated in a
    neighborhood of the phase transition point.  The curves shown here
    include the original GM1L equation of state, $k=0$, and several
    equations of state with larger density offsets, $0< k \leq
    100$. The $k=100$ curve has the maximum density offset allowed in
    stable (and therefore observable) neutron-stars.}
\end{figure}

The particular parametric equation of state representations used in
this study~\cite{Lindblom2010, Lindblom2022} are described in
Appendix~\ref{s:ParametricRepresentations}.  These parametric
representations are causal in the sense that the sound speeds are less
than the speed of light for every choice of the parameters.  The
representations used in this study express the energy density
$\epsilon(h,\upsilon_a)$ and the pressure $p(h,\upsilon_a)$ as
functions of the enthalpy $h$ of the fluid and a set of parameters
$\upsilon_a$ for $1\leq a\leq N_\mathrm{parms}$.  This type of
representation is most useful when using the enthalpy-based form of
the relativistic stellar structure equations~\cite{Lindblom1992}. This
form of the equations allows numerical determinations of the masses
and radii more accurately and more efficiently than the standard
pressure-based Oppenheimer-Volkoff~\cite{Oppenheimer1939} form. The
families of exemplar equations of state with phase transitions
described in Appendix~\ref{s:ExemplarEOS} were produced as
enthalpy-based tables, $\{\epsilon_i,p_i,h_i\}$ for $0\leq i \leq
N_\mathrm{table}$, to facilitate comparisons with the enthalpy-based
parametric representations.

\begin{figure}[!t]
  \centerline{
    \includegraphics[width=0.4\textwidth]{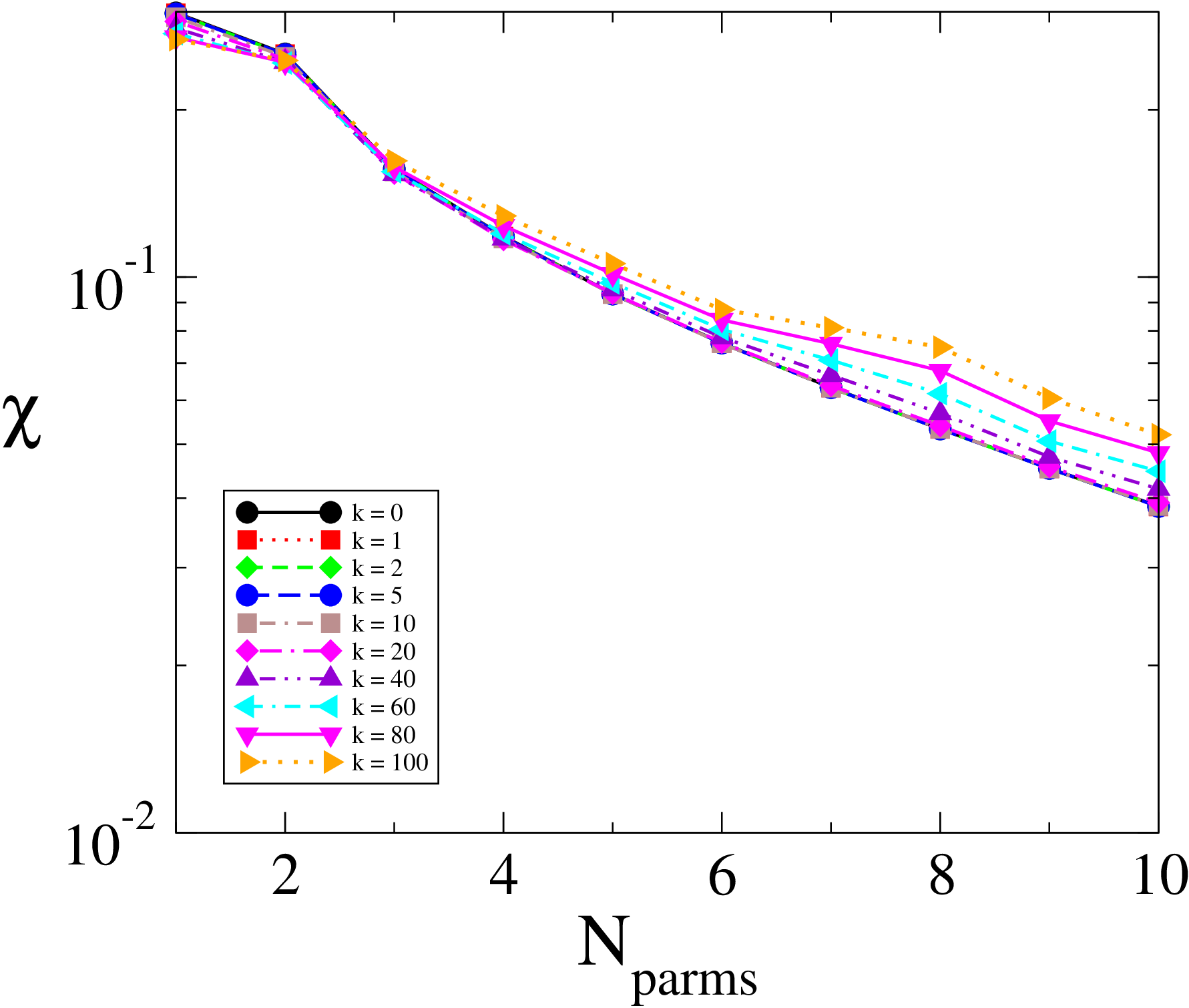}}
  \caption{\label{f:FOOptimalFits_PA}Average errors
    $\chi(N_\mathrm{parms})$ for the piecewise-analytic fits as a
    function of $N_\mathrm{parms}$ for a family of equations of state
    with first-order phase transitions of various sizes, $0\leq k \leq
    100$.}
\end{figure}
\begin{figure}[!h]
  \centerline{
    \includegraphics[width=0.4\textwidth]{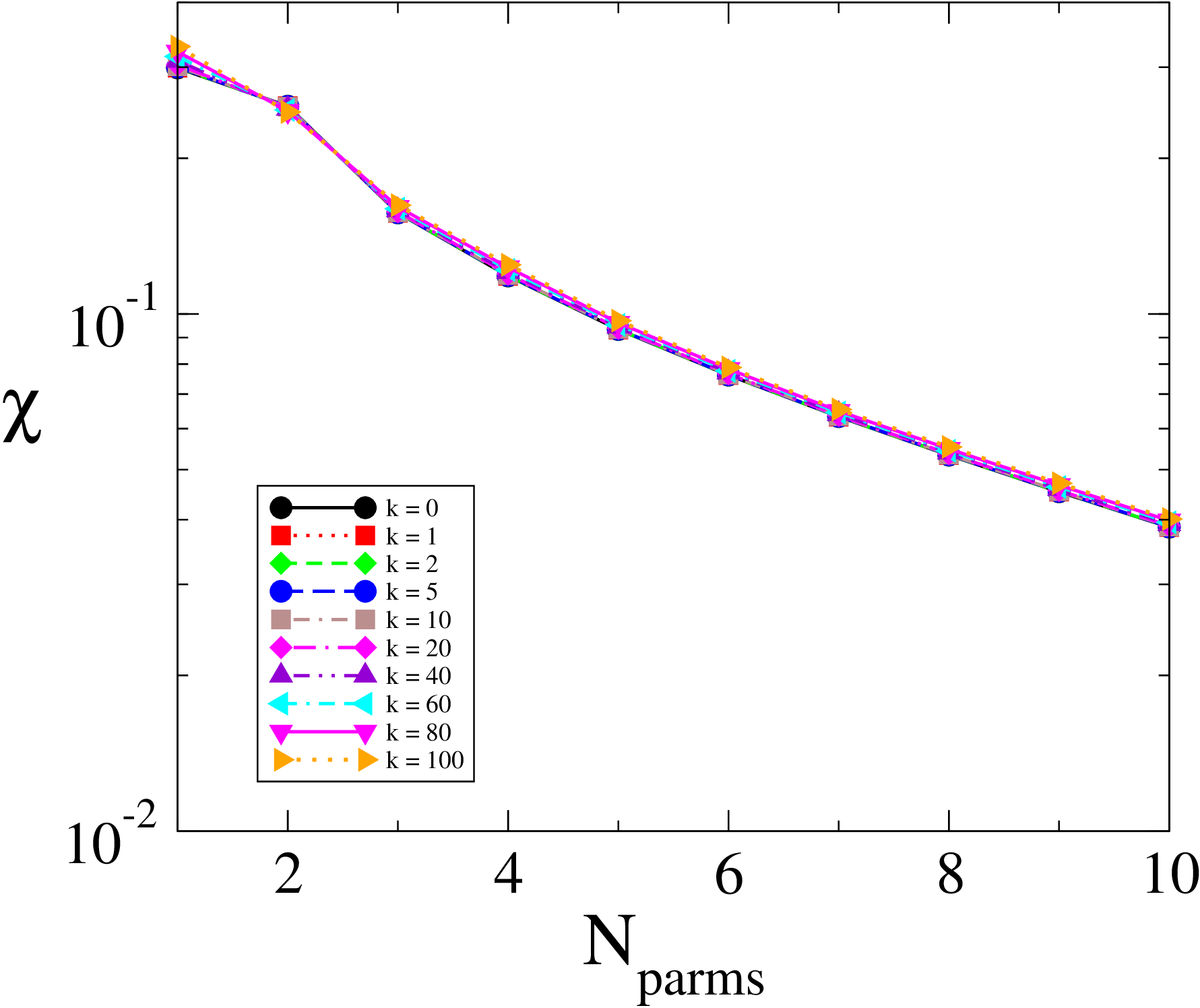}}
  \caption{\label{f:SOOptimalFits_PA}Average errors
    $\chi(N_\mathrm{parms})$ for the piecewise-analytic fits as a
    function of $N_\mathrm{parms}$ for a family of equations of state
    with second-order phase transitions of various sizes, $0\leq k \leq
    100$.}
\end{figure}
\begin{figure}[!h]
  \centerline{
    \includegraphics[width=0.4\textwidth]{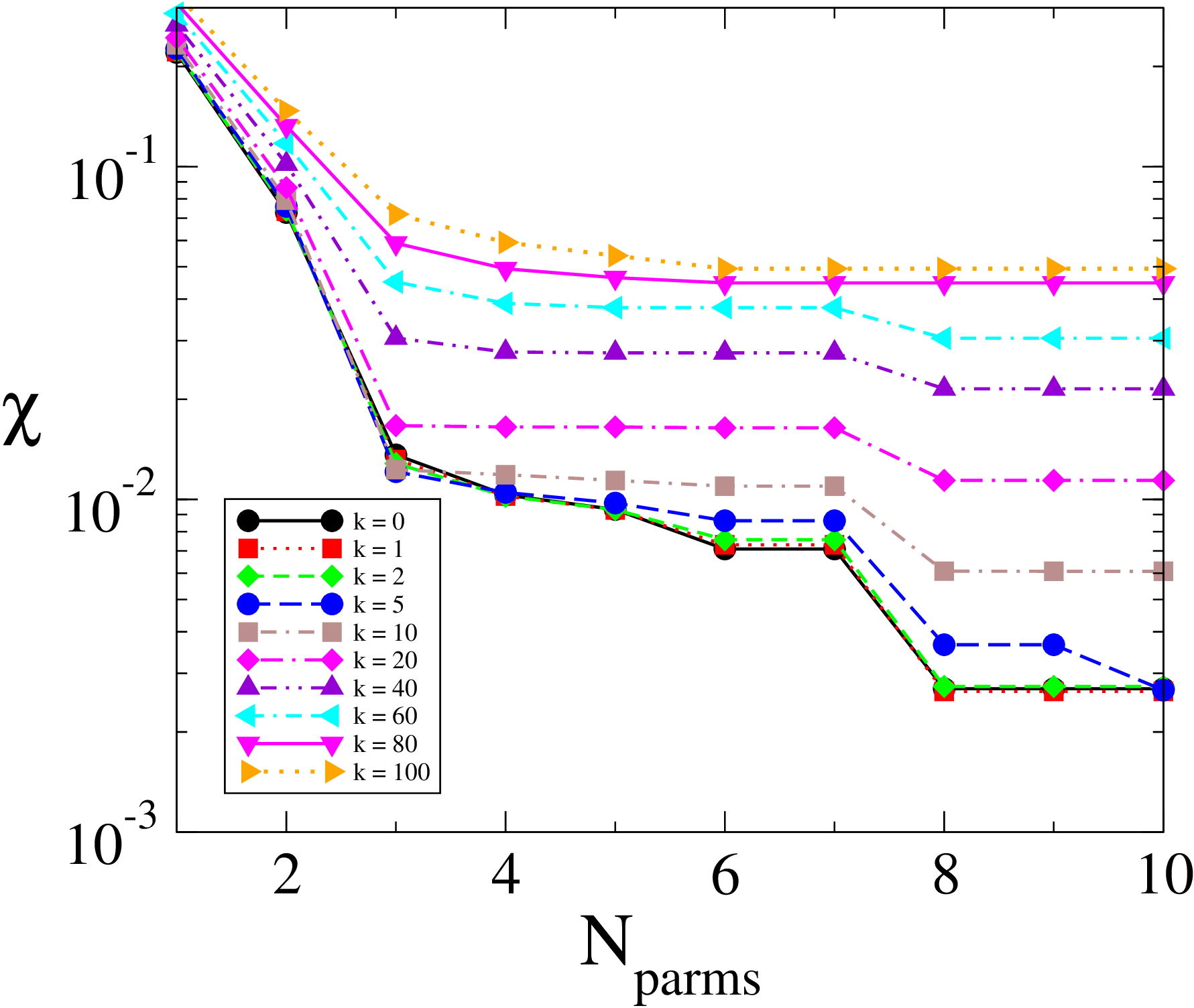}}
  \caption{\label{f:FOOptimalFits}Average errors $\chi$ for the
    optimal spectral fits as a function of $N_\mathrm{parms}$ for a
    sequence of equations of state with first-order phase transitions
    of various sizes, $0\leq k \leq 100$.}
\end{figure}
\begin{figure}[!h]
  \centerline{
    \includegraphics[width=0.4\textwidth]{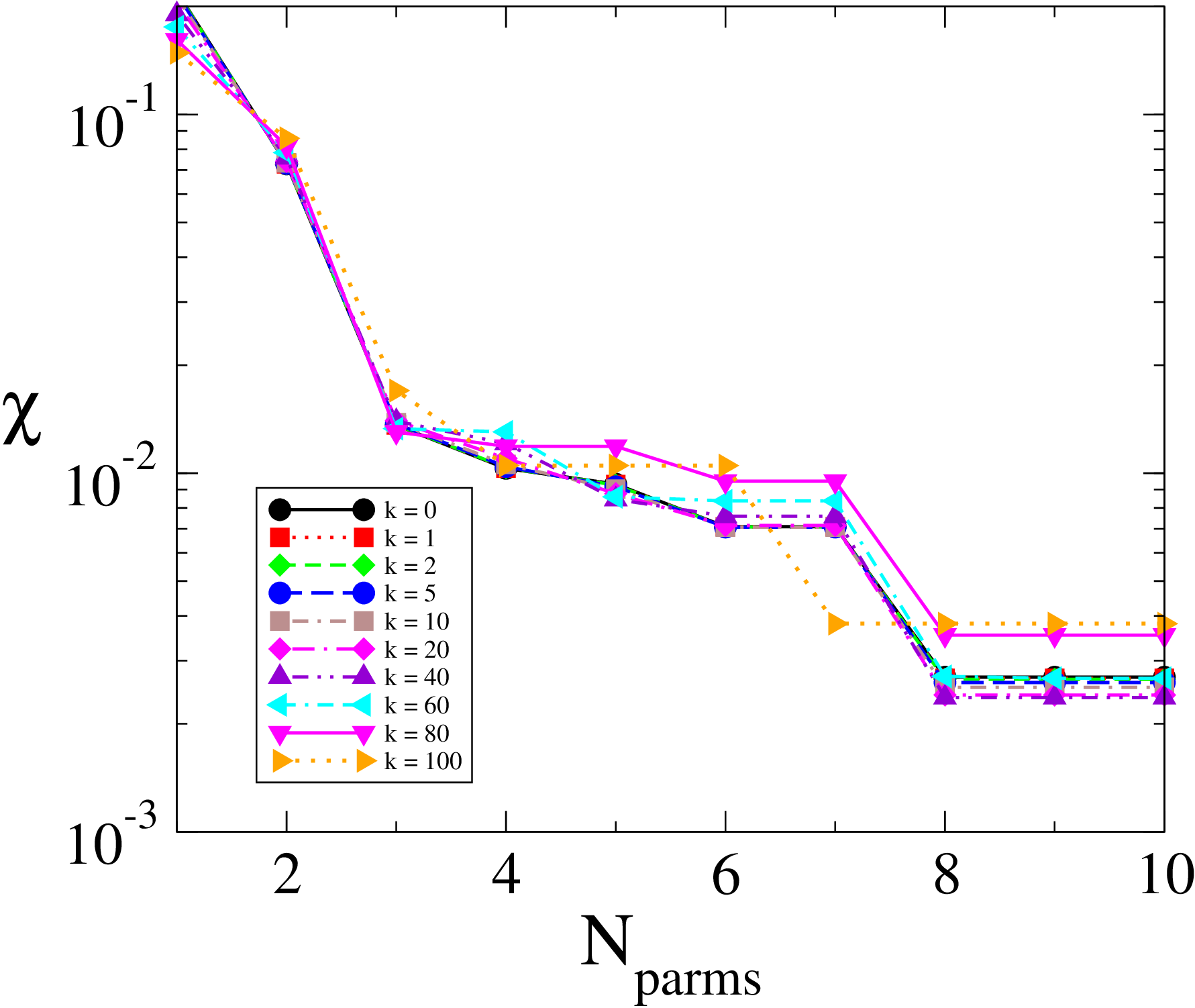}}
  \caption{\label{f:SOOptimalFits}Average errors $\chi$ for the
    optimal spectral fits as a function of $N_\mathrm{parms}$ for a
    sequence of equations of state with second-order phase transitions
    of various sizes, $0\leq k \leq 100$.}
\end{figure}
The next step is to find the values of the parameters $\upsilon_a$ in
the equation of state, $\epsilon(h,\upsilon_a)$, that best
approximates one of the exemplar equations of state.  The optimal
parameter values $\upsilon_a$ are found in this study by minimizing
the function $\chi(\upsilon_a)$ that measures the average difference
between the tabulated values of the exemplar equation of state,
$\epsilon_i(h_i)$, and the corresponding values from the parametric
equation of state, $\epsilon(h_i,\upsilon_a)$:
\begin{equation}
  \chi^2(\upsilon_a)=\frac{1}{N_\mathrm{table}}\sum_{i=0}^{N_\mathrm{table}}
  \left[\log\left(\frac{\epsilon(h_i,\upsilon_a)}{\epsilon_i(h_i)}
    \right)\right]^2.
  \label{e:chisqr_Def}
\end{equation}
The error function $\chi^2(\upsilon_a)$ is non-negative, and therefore
has a minimum for some $\upsilon_a$.  The minimization of
$\chi(\upsilon_a)$ is carried out numerically in this study using an
algorithm based on the Levenberg-Marquardt method~\cite{numrec_f}.
The equation of state, $\epsilon(h,\upsilon_a)$ and $p(h,\upsilon_a)$,
produced by this minimization process is the optimal parametric fit to
this equation of state.

Model equations of state created with different numbers of parameters,
$N_\mathrm{parms}$, produce different error minima,
$\chi^2(\upsilon_a,N_\mathrm{parms})$.  Those with larger
$N_\mathrm{parms}$ generally produce smaller errors, and therefore
provide better approximations to the original tabulated equation of
state.  The minimum values of $\chi(N_\mathrm{parms})$ for the causal
piecewise-analytic representations of the exemplar equations of state
with first-order phase transitions are shown as functions of
$N_\mathrm{parms}$ in Fig.~\ref{f:FOOptimalFits_PA} for a range of
discontinuity sizes, $0\leq k \leq 100$.
Figure~\ref{f:SOOptimalFits_PA} shows the analogous results for the
exemplar equations of state with second-order phase transitions.  The
results for the causal spectral representations of the exemplar
equations of state with first- or second-order phase transitions are
shown in Figs.~\ref{f:FOOptimalFits} and \ref{f:SOOptimalFits}
respectively.

\section{Discussion}
\label{s:Discussion}

The results in Figs.~\ref{f:FOOptimalFits_PA} and
\ref{f:SOOptimalFits_PA} show that the piecewise-analytic parametric
fits to the exemplar equations of state are convergent, in the sense
that the average errors $\chi(N_\mathrm{parms})$ decrease
monotonically as the number of parameters $N_\mathrm{parms}$
increases.  These results also show that the accuracies of the
piecewise-analytic representations do not depend strongly on the size
of the discontinuities.  The piecewise-analytic representations
therefore provide a robust way to represent equations of state with
discontinuities from first- or second-order phase transitions.

The results in Figs.~\ref{f:FOOptimalFits} and \ref{f:SOOptimalFits}
for the spectral representations are more nuanced.  The modeling
errors $\chi(N_\mathrm{parms})$ decrease rapidly as $N_\mathrm{parms}$
increases to $N_\mathrm{parms}=8$ for equations of state with small
discontinuities.  However for larger values, $N_\mathrm{parms}>8$, and
for equations of state with larger discontinuities,
$\chi(N_\mathrm{parms})$ becomes more or less constant.  These results
show that the particular spectral representation used in this study
does not provide convergent representations of equations of state with
discontinuities caused by phase transitions.

\begin{figure}[!t] 
  \centerline{
    \includegraphics[width=0.4\textwidth]
               {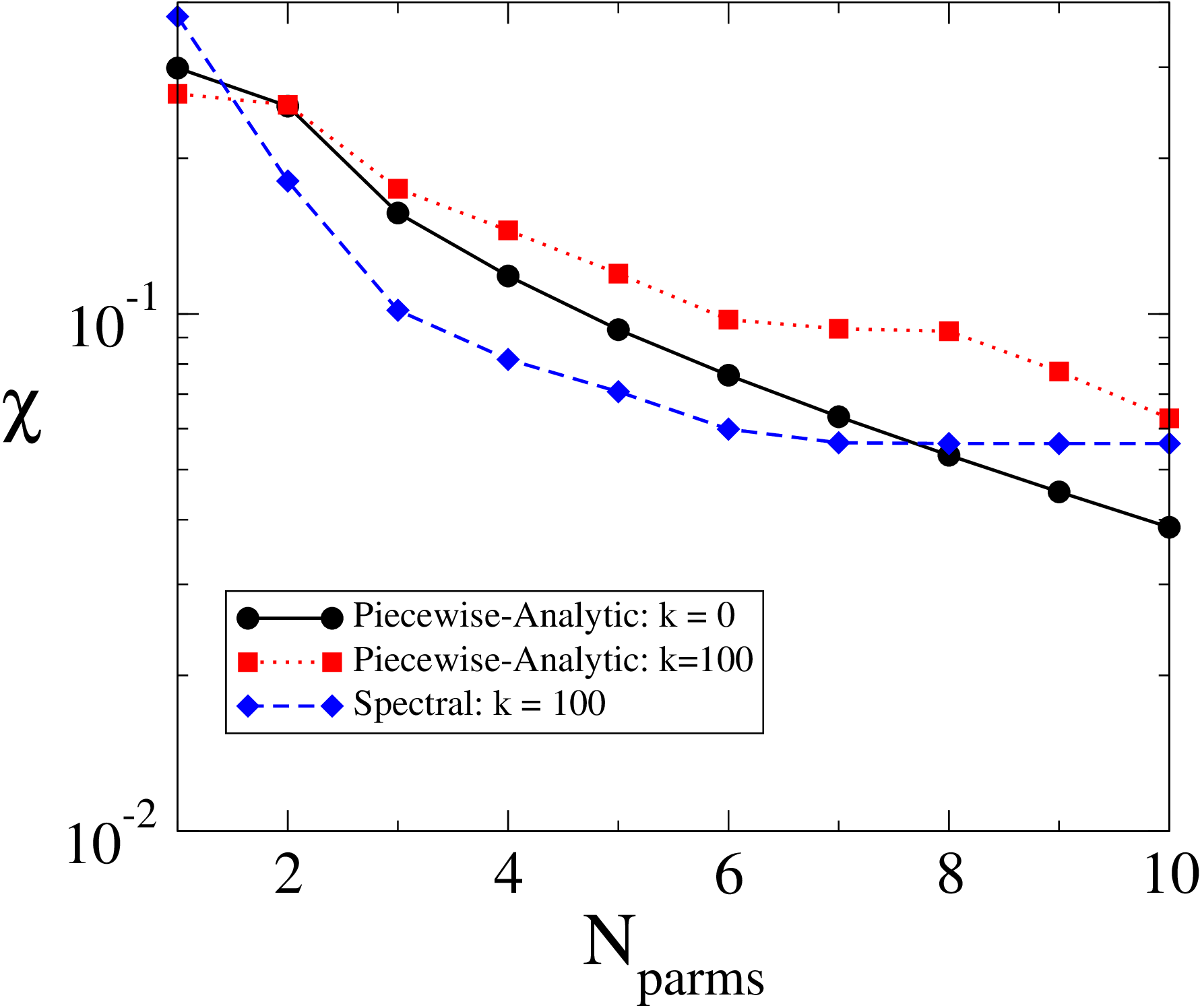}}
  \caption{\label{f:FO_RelativeConvergence}Comparing modeling errors,
    $\chi(N_\mathrm{parms})$, for the optimal causal
    piecewise-analytic and the causal spectral fits to the exemplar
    neutron-star equations of state with first-order phase
    transitions.}
\end{figure}

\begin{figure}[!t] 
  \centerline{
    \includegraphics[width=0.4\textwidth]{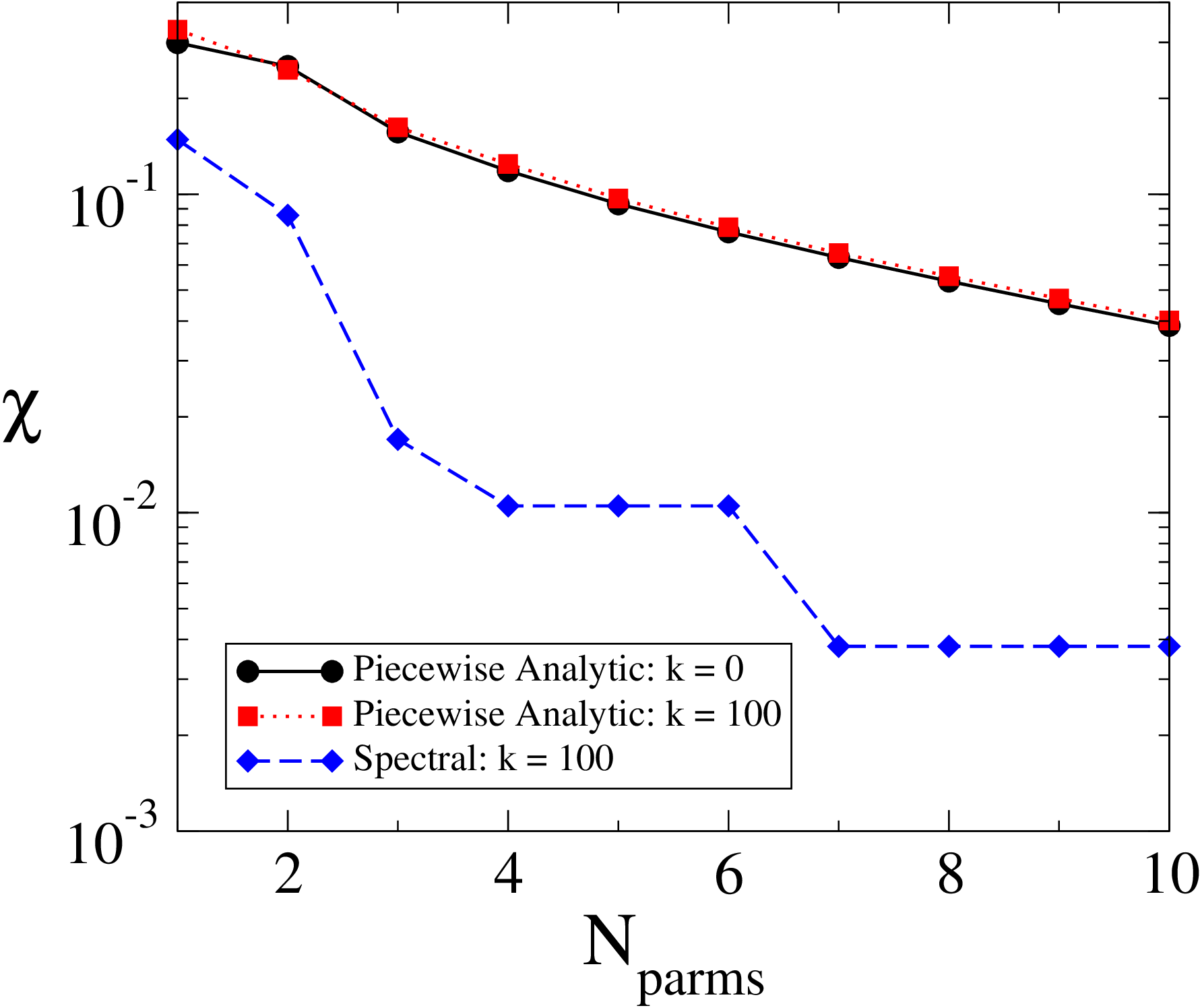}}
  \caption{\label{f:SO_RelativeConvergence}Comparing modeling errors,
    $\chi(N_\mathrm{parms})$, for the optimal causal
    piecewise-analytic and the causal spectral fits to the exemplar
    neutron-star equations of state with second-order phase
    transitions.}
\end{figure}

Nevertheless, the modeling errors $\chi(N_\mathrm{parms})$ for the
low-order, $2\leq N_\mathrm{parms}\leq 7$, spectral representations
are smaller than those of the corresponding piecewise-analytic
representations for every exemplar equation of state included in this
study.  Figures~\ref{f:FO_RelativeConvergence} and
\ref{f:SO_RelativeConvergence} illustrate the relative accuracies of
the two types of parametric representation for the equations of state
with first- or second-order phase transitions respectively.  While
these spectral representations are not convergent, these results show
that they are still the most accurate choice when using low-order
parametric fits.  The errors in the $N_\mathrm{parms}=3$ spectral
fits, for example, are fairly small, $0.012 \leq \chi\leq 0.072$, for
all the phase transitions studied here.  Until the quality and
quantity of observational data are improved to allow more accurate
determinations of the equation of state, the low-order spectral
representations are likely to be the best choice.

It is not clear why the spectral representations fail to converge for
equations of states with discontinuities.  The basis functions used in
the particular spectral representation used here are simple powers of
$\log\left(h/h_0\right)$, see Eq.~(\ref{e:Upsilon_spectral}).  This
spectral expansion is therefore similar in form to a Taylor expansion
of the velocity function.  The radius of convergence of the Taylor
expansion of this function would not extend into the high density
region beyond the discontinuity caused by a phase transition.  It is
possible that the spectral expansions using these power-law spectral
basis functions fail to converge for a similar reason.  If this is the
problem, then changing basis functions to Chebyshev polynomials or
Fourier basis functions whose domains span the phase transition point
would likely improve the convergence properties of the spectral
representations.

In cases where a strong first-order phase transition is present,
better accuracy and numerical convergence could also be achieved by
dividing the equation of state into a low-density domain with
pressures and energy densities below the phase transition point, and a
second high-density domain with pressures and densities above that
point.  In the low-density domain, $h_0\leq h \leq h_{\,T}$, the
spectral expansion defined in
Eqs.~(\ref{e:p_h_integral})--(\ref{e:SpectralExpansion}) could be
used.  In the high-density domain a separate but similar spectral
expansion could be used:
\begin{eqnarray}
  p(h) &=& p_T + (\bar \epsilon_T c^2+p_T)\int_{h_T}^h\bar \mu(h')\,dh',
  \label{e:p_h_integral_bar}\\
  \epsilon(h) &=& -\frac{p(h)}{c^{2}}
  + \left(\bar \epsilon_T+\frac{p_T}{c^{2}}\right)\,\bar\mu(h),
  \label{e:epsilon_h_integral_bar}
\end{eqnarray}
where $p_T=p(h_T)$ and $\bar \epsilon_T$ represent the point on the
equation of state curve just above the phase transition point. The
quantity $\bar \mu(h)$ used in these expressions is given by
\begin{equation}
  \bar \mu(h)= \exp\left\{
  \int_{h_T}^h\left[2+\bar\Upsilon(h')\right]\,dh'\right\},
  \label{e:mu_h_integral_bar}
\end{equation}
while the velocity function $\bar \Upsilon(h)$ used in the
high-density domain is given by,
\begin{equation}
  \bar \Upsilon(h,\upsilon_a)=\exp\left\{\,
  \sum_{a=1}^{\bar N_\mathrm{parms}}\,\bar \upsilon_a
  \Phi_a(h)\right\},
  \label{e:Upsilon_spectral_bar}
\end{equation}
for some suitable choice of basis functions $\Phi_a(h)$.

The single domain spectral expansions considered in this study are
defined by the values of the $N_\mathrm{parms}$ spectral parameters
$\upsilon_a$.  The two-domain spectral expansions are defined by the
values of the original $N_\mathrm{parms}$ spectral parameters
$\upsilon_a$, plus the $\bar N_\mathrm{parms}$ spectral parameters
$\bar \upsilon_a$, plus two additional parameters $h_{\,T}$ and
$\bar\epsilon_T$ that determine where the phase transition is located
and the size of the energy-density discontinuity at that point.  Thus
the number of parameters needed to specify the equation of state using
this two-domain approach grows from $N_\mathrm{parms}$ to
$N_\mathrm{parms}+\bar N_\mathrm{parms}+2$.  The cost of going to a
two-domain spectral representation could only be justified if
sufficient accuracy could not be achieved using a single-domain
representation with the same total number of parameters.

\appendix

\section{Exemplar Equations of State}
\label{s:ExemplarEOS}

The exemplar equations of state used in this study were constructed by
introducing discontinuities into the relatively smooth GM1L equation
of state, which is based on a mean-field representation of the
interactions between nucleons~~\cite{Typel2010}.  The basic
representation of GM1L used here is a table of energy density and
pressure points: $\{\epsilon_i,p_i\}$ for $0\leq i \leq
N_\mathrm{table}$.  The primary goal of this study is to test the
accuracy of parametric representations of equations of state with
phase transitions in the nuclear-density range.  Consequently only the
high-density portion of the GM1L equation of state table is used here,
beginning at the table entry, $\{\epsilon_0,p_0\}$, where
$\epsilon_0=5.08587\times 10^{13}$ g/cm${}^3$ and $p_0=1.20788\times
10^{32}$ erg/cm${}^3$.

This study uses enthalpy-based representations of the equation of
state, so the basic pressure-based GM1L table, $\{\epsilon_i,p_i\}$,
must be converted to an enthalpy-based table:
$\{\epsilon_i,p_i,h_i\}$.  The enthalpy of a relativistic fluid is
defined by,
\begin{equation}
  h(p)=h_0+\int_{p_0}^{p'}\frac{dp'}{\epsilon(p')c^2+p'}.
  \label{e:Enthalpy_Def}
\end{equation}
In order to evaluate this integral for the tabulated GM1L equation of
state, an interpolation formula must be used to determine the values
of $\epsilon(p)$ between table entries.  The commonly used
pseudo-polytropic interpolation,
\begin{equation}
  p = p_i \left(\frac{\epsilon}{\epsilon_i}\right)^{\Gamma_i},
  \label{e:EOS_interpolation}
\end{equation}
is used here to define this equation of state for energy densities,
$\epsilon_i\leq \epsilon < \epsilon_{i+1}$, in the intervals between
table entries.  The constants $\Gamma_i$ in this expression are
defined by
\begin{equation}
  \Gamma_i=\frac{\log(p_{i+1}/p_i)}{\log(\epsilon_{i+1}/\epsilon_i)}.
  \label{e:GammaDef}
\end{equation}
The low density value of $h_0=p(h_0)$ used in this study is determined
by evaluating the enthalpy integral in the low-density range using one
of the standard lower density neutron-star equations of
state~\cite{BPS1971}, with the result $h_0\approx 1.74067\times
10^{-2}$.  At higher densities the enthalpy can be determined by
integrating Eq.~(\ref{e:Enthalpy_Def}) between table entries $p_i$ and
$p_{i+1}$ using Eq.~(\ref{e:EOS_interpolation}).  These integrals can
be done analytically resulting in a recursion
relation for the $h_{i+1}$ table entries~\cite{Lindblom2012}:
\begin{equation}
  h_{i+1}=h_i+\frac{\Gamma_i}{\Gamma_i-1}
  \log\left[\frac{\epsilon_{i}(\epsilon_{i+1}\, c^2 +p_{i+1})}
    {\epsilon_{i+1}(\epsilon_i\, c^2 +p_i)}\right].
  \label{e:enthalpy_recursion}
\end{equation}

Discontinuities are inserted into GM1L for this study at a point
several times nuclear density where the energy density has the value
$\epsilon_{\,T}=8\times 10^{14}$ g/cm${}^3$.  The particular equations
of state with discontinuities representing first- or second-order
phase transitions are described in Secs.~\ref{s:FirstOrder} and
\ref{s:SecondOrder} respectively.  These exemplar equations of state
have discontinuities with a range of sizes from zero to the largest
physically relevant phase transition of each type.

\subsection{First-Order Phase Transitions}
\label{s:FirstOrder}

An equation of state $\epsilon=\epsilon(p)$ is said to have a
first-order phase transition at $p=p_{\,T}$ if $\epsilon(p)$ is
discontinuous at that point.  Exemplar equations of state with
first-order phase transitions are constructed here by modifying the
GM1L equation of state at densities above $\epsilon_{\,T}$. For this
study the transition density $\epsilon_{\,T}$ is chosen to be several
times nuclear density at the point $\epsilon_{\,T}=8\times 10^{14}$ g/cm${}^3$.

To ensure the tabulated representations of the exemplar equations of
state adequately represent the sharp transitions at the phase
transition, points are added to the table entries at the points
$\epsilon_{\,T}^\pm=(1\pm 10^{-6})\,\epsilon_\mathrm{\,T}$, just above
and below the phase transition.  The corresponding pressure points
needed to complete the table entries are given by
Eq.~(\ref{e:EOS_interpolation}): $p^\pm_\mathrm{\,T}=
p_i\,(\epsilon^\pm_\mathrm{\,T}/\epsilon_i)^{\Gamma_i}$, where
$\epsilon_i<\epsilon^\pm_{\,T} < \epsilon_{i+1}$.  Once the GM1L
equation of state table has been updated with these two additional
phase-transition bracketing points, density offsets
$\delta\epsilon_{\,T}$ are added to all the table entries with
densities above $\epsilon_{\,T}$.  The result is a tabulated model
equation of state with a first-order phase transition.

The neutron-star mass-radius curves produced by equations of state
with first-order phase transitions show that stars with central
densities above $\epsilon_{\,T}$ are unstable whenever the density
discontinuity $\delta\epsilon_{\,T}$ exceeds a certain maximum,
$\max(\delta\epsilon_{\,T})$~\cite{Lindblom98a}.  The masses of these
stars achieve a maximum at the point where the central density equals
$\epsilon_{\,T}$.  Stars with larger central densities are subject to
a gravitational instability predicted by general relativity theory.
The family of stable neutron stars therefore terminates at this point.
In some cases there may be an additional higher density family of
stable ``hyperon'' or perhaps ``quark'' stars.\footnote{If the
  mass-radius curve of these stars has a second inflection point
  beyond $\{\epsilon_{\,T},p_{\,T}\}$ where the mass has a minimum and
  the radius is decreasing, then stability would be restored and a
  higher density branch of relativistic stars could exist.}  Neutron
stars with central pressures in the unstable region above
$\epsilon_{\,T}$ can never be observed, so equations of state with
density offsets above $\max(\delta\epsilon_{\,T})$ will not be
considered in this study.  The approximate value of this maximum
density offset is given by~\cite{Lindblom98a},
\begin{equation}
  \max(\delta\epsilon_{\,T})=\frac{1}{3}
  \left(\epsilon_{\,T}+3\,\frac{p_{\,T}}{c^2}\right),
  \label{e:MaxDensityOffset}
\end{equation}
where $p_\mathrm{\,T}=
p_i\,(\epsilon_\mathrm{\,T}/\epsilon_i)^{\Gamma_i}$, and
$\epsilon_i<\epsilon_{\,T} < \epsilon_{i+1}$.  The maximum density
offset, $\max(\delta\epsilon_{\,T})$, is fairly large for the model
first-order phase transitions constructed here:
$\max(\delta\epsilon_{\,T})/ \epsilon_{\,T}\approx
0.504634$.\footnote{The maximum density offset derived by an
analytical analysis in Ref.~\cite{Lindblom98a} is 1.5 times the value
given in Eq.~(\ref{e:MaxDensityOffset}). Numerical studies, however,
show that the effective maximum offset is close to the value given in
Eq.~(\ref{e:MaxDensityOffset}).}
  
\begin{figure}[!h]
  \centerline{
    \includegraphics[width=0.4\textwidth]{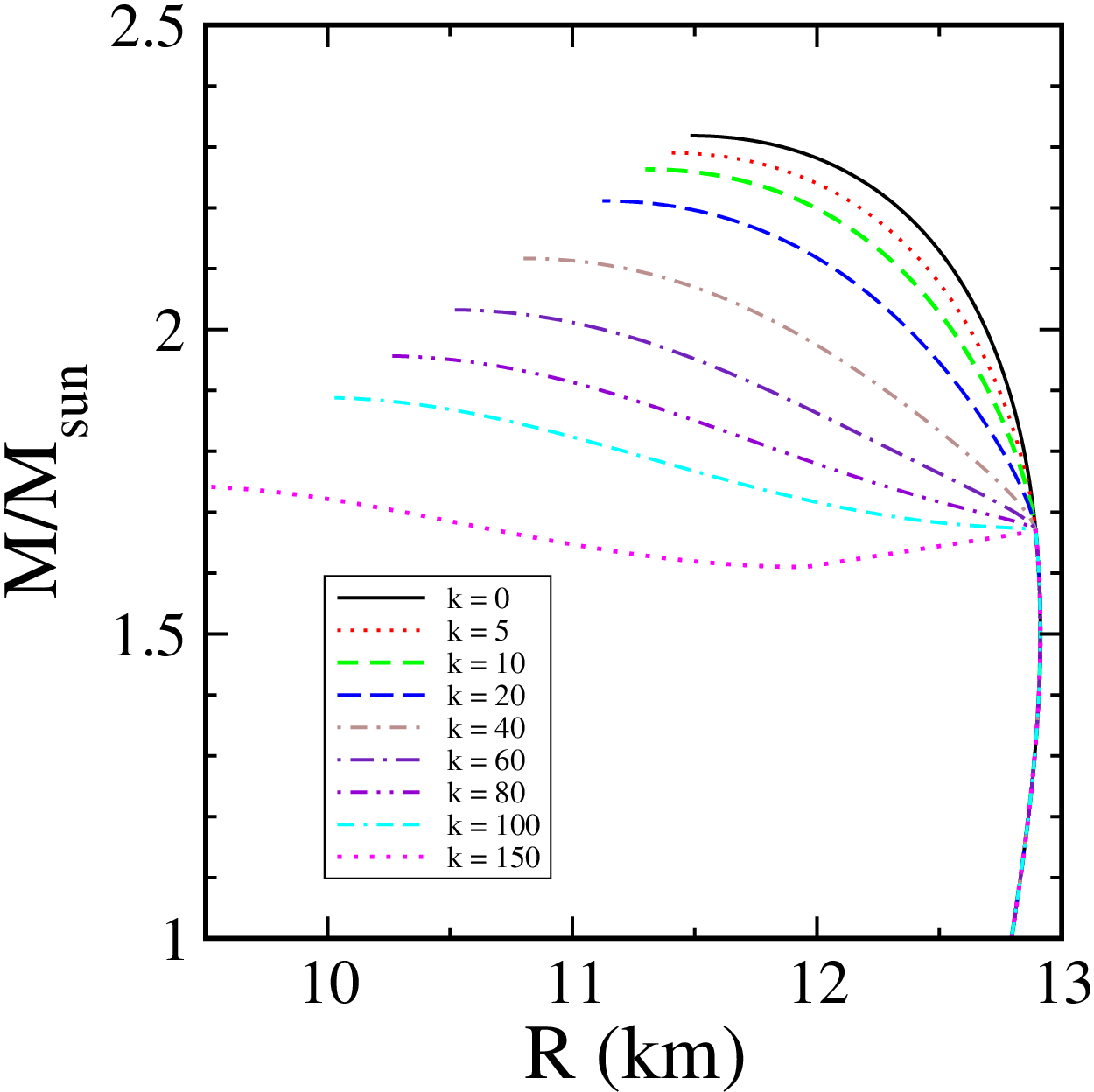}}
  \caption{\label{f:FO_MofR}Examples of mass-radius curves for
    equations of state with first-order phase transitions.  The $k=0$
    curve is based on the unmodified GM1L equation of state, while the
    $k=100$ curve corresponds to the equation of state with the
    maximum density discontinuity defined in
    Eq.~(\ref{e:MaxDensityOffset}). The $k=150$ curve illustrates an
    equation of state with a larger density discontinuity which leads
    to unstable stars beyond the $\{\epsilon_{\,T},p_{\,T}\}$ phase
    transition point.  This $k=150$ curve also has a disconnected
    branch of stable relativistic stars at higher densities.}
\end{figure}

A family of exemplar equations of state with first-order phase
transitions have been constructed for this study with density offsets
$\delta\epsilon_{\,T}$ having sizes in the range $0\leq
\delta\epsilon_{\,T}\leq \max(\delta\epsilon_{\,T})$.  The density
offsets used in these models are given by,
\begin{equation}
  \delta\epsilon_{\,T}=\sigma_k\,\max(\delta\epsilon_{\,T}),
\end{equation}
where the size of the offsets is determined by
\begin{equation}
  \sigma_k=\frac{k}{100},
  \label{e:mu_k_def}
\end{equation}
for $0\leq k \leq 100$.  The density offset for each exemplar equation
of state is added to all the table entries that exceed the transition
density $\epsilon_{\,T}$.  Figure~\ref{f:FirstOrderExamples}
illustrates some of these exemplar equations of state in a
neighborhood of the phase transition point.  Figure~\ref{f:FO_MofR}
illustrates the mass-radius curves generated from these exemplar
equations of state with first-order phase transitions.

\subsection{Second-Order Phase Transitions}
\label{s:SecondOrder}

Second-order phase transitions are points on the equation of state
curve $\epsilon=\epsilon(p)$ where $\epsilon(p)$ is continuous but
$d\epsilon/dp$ is discontinuous.  In this study the derivatives of the
GM1L equation of state are modified at and above the phase transition
density $\epsilon_{\,T}$ to create a discontinuity that simulates a
second-order phase transition.  To do this most efficiently, the basic
GM1L equation of state table is modified by inserting an additional
entry at the point $\{\epsilon_{\,T},p_{\,T}\}$.  With this addition
the second-order phase-transition discontinuity in $d\epsilon/dp$
occurs at one of the tabulated data points.

The derivative of the equation of state, $d\epsilon/dp$, is related to
the speed of sound $v$ in a barotropic fluid by
$v=(d\epsilon/dp)^{-1/2}$.  Any modifications of $d\epsilon/dp$ above
the phase transition point must therefore be done in a way that
respects causality.  A convenient tool for monitoring the causality of
sound waves in fluids is the dimensionless velocity function
$\Upsilon$ defined by~\cite{Lindblom2018}
\begin{equation}
  \Upsilon  = c^2\frac{d\epsilon}{dp} - 1 = \frac{c^2 - v^2}{v^2}.
  \label{e:UpsilonDef}
\end{equation}
The propagation of sound waves is causal if and only if $\Upsilon\geq
0$.  The velocity function $\Upsilon$ is determined from the basic
tabulated GM1L equation of state by evaluating $d\epsilon/dp$ using
the interpolation formula in Eq.~(\ref{e:EOS_interpolation}) at each
point in the table.  The result is given by
\begin{equation}
  \Upsilon_i=\frac{\epsilon_i\,c^2}{p_i\Gamma_i}-1.
  \label{e:Upsilon_i}
\end{equation}
For causal equations of state $\Upsilon\geq 0$, with $\Upsilon=0$
representing the extreme case of a fluid with sound speed equal to the
speed of light, $v^2=c^2$.

Discontinuities in the slope of the exemplar equations of state were
introduced for this study by modifying $\Upsilon$ for densities above
the phase transition density, $\epsilon\geq \epsilon_\mathrm{\,T}$,
while leaving it unchanged for lower densities. In particular the
sound speed was increased by reducing $\Upsilon$ by multiplying it by
the factor $1-\sigma_k$ in this high density region, using the $\sigma_k$
defined in Eq.~(\ref{e:mu_k_def}).  Thus $\Upsilon$ is replaced in
this high density region by $\tilde\Upsilon$ defined by
\begin{equation}
  \tilde \Upsilon = (1-\sigma_k)\,\Upsilon.
  \label{e:tildeUpsilonDef}
\end{equation}

The maximum physically relevant slope discontinuity at the phase
transition point is achieved by setting the sound speed to the speed
of light, $v^2=c^2$ at that point, i.e. by setting $\tilde\Upsilon=0$
there.  A family of exemplar equations of state models were
constructed for this study that range from the original undistorted
GM1L equation of state for $k=0$, to the extreme equation of state
with $\tilde\Upsilon=0$ above the transition density for $k=100$.

\begin{figure}[!h]
  \centerline{
    \includegraphics[width=0.4\textwidth]{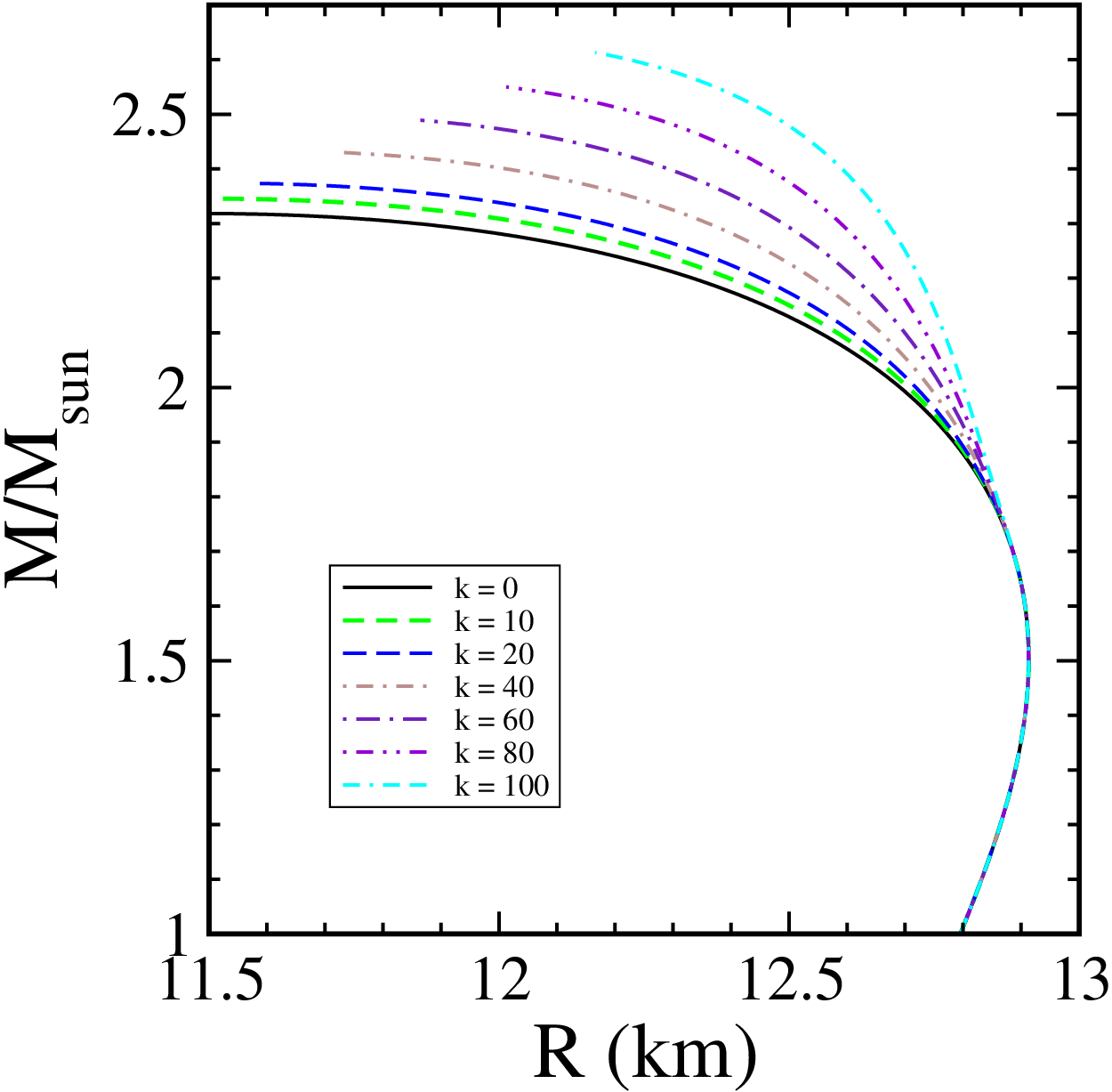}}
  \caption{\label{f:SO_MofR}Examples of mass-radius curves for
    equations of state with second-order phase transitions.  The $k=0$
    curve is based on the unmodified GM1L equation of state, while the
    $k=100$ curve is based on the maximal case where the sound speed
    changes discontinuously to the speed of light above the phase
    transition.}
\end{figure}

Given $\tilde\Upsilon_i$ evaluated at the points of the basic GM1L
equation of state table, the modified values of $\epsilon_i$ above the
phase transition point, $\epsilon_i>\epsilon_{\,T}$, can be determined
by the recursion relation
\begin{equation}
  \epsilon_{i+1}=\epsilon_i
  \exp\left[(1+\tilde\Upsilon_i)\frac{p_i}{\epsilon_i\,c^2}
    \log\left(\frac{p_{i+1}}{p_i}\right)\right].
\end{equation}
This expression follows by solving Eq.~(\ref{e:Upsilon_i}) for
$\epsilon_{i+1}$ which contributes to the definition of $\Gamma_i$.
The pressure points $p_i$ in the equation of state table are not
modified.  Figure~\ref{f:SecondOrderExamples} illustrates a few of
these exemplar equations of state with larger density-derivative
discontinuities in a neighborhood of the second-order phase transition
point. Figure~\ref{f:SO_MofR} illustrates the mass-radius curves
generated from these exemplar equations of state with second-order
phase transitions.

\section{Causal Parametric Representations}
\label{s:ParametricRepresentations}

This study uses enthalpy-based representations of the neutron-star
equation of state.  These representations determine the energy density
$\epsilon(h,\upsilon_a)$ and pressure $p(h,\upsilon_a)$ as functions
of the enthalpy $h$ and a collection of $N_\mathrm{parms}$ parameters
$\upsilon_a$ for $1\leq a\leq N_\mathrm{parms}$.  To be useful tools
for representing the physical neutron-star equation of state, these
representations must be faithful and they must be causal.  Faithful
representations have the property that every choice of parameters,
$\upsilon_a$, represents a possible physical equation of state.
Conversely every physical equation of state can be represented by some
choice, including perhaps an infinite sequence, of
parameters~\cite{Lindblom2010}. Faithful parametric representations
must be convergent as the number of parameters is increased. Causal
representations have the property that every choice of parameters
generates an equation of state with sound speeds less than or equal to
the speed of light~\cite{Lindblom2018}.

An equation of state has causal sound speeds if and only if the
velocity function $\Upsilon$, defined in Eq.~(\ref{e:UpsilonDef}), is
non-negative: $\Upsilon\geq 0$.  This velocity function can be used as
a generating function that determines the full equation of state, so
it is a very useful tool for constructing causal parametric
representations.  The velocity function can be written as a function
of the enthalpy:
\begin{equation}
  \Upsilon(h)=c^2\frac{d\epsilon}{dp}-1 =
  c^2\frac{d\epsilon}{dh}\,\left[\epsilon(h)\, c^2+p(h)\right]^{-1}-1.
  \label{e:Upsilon_h_Def}
\end{equation}
Given a velocity function, $\Upsilon(h)$, the full equation of state
can be reconstructed by solving the following ordinary differential
equations for $\epsilon(h)$ and $p(h)$,
\begin{eqnarray}
  \frac{dp}{dh} &=& \epsilon\, c^2+p. 
  \label{e:p_h_ode}\\
  \frac{d\epsilon}{dh}&=&
  \left(\epsilon+\frac{p}{c^2}\right)\left[\Upsilon(h)+1\right],
  \label{e:epsilon_h_ode}
\end{eqnarray}
The first, Eq.~(\ref{e:p_h_ode}), follows from the definition of the
enthalpy in Eq.~(\ref{e:Enthalpy_Def}), while the second,
Eq.~(\ref{e:epsilon_h_ode}), follows from the definition of
$\Upsilon(h)$ in Eq.~(\ref{e:Upsilon_h_Def}).  These equations can be
reduced to quadratures:
\begin{eqnarray}
  p(h) &=& p_0 + (\epsilon_0\, c^2 +p_0)\int_{h_0}^h\mu(h')\,dh',
  \label{e:p_h_integral}\\
  \epsilon(h) &=& -p(h) + \left(\epsilon_0+\frac{p_0}{c^2}\right)\,\mu(h),
  \label{e:epsilon_h_integral}
\end{eqnarray}
where $p_0=p(h_0)$ and $\epsilon_0=\epsilon(h_0)$ represent a point on
the equation of state curve, and $\mu(h)$ is given by
\begin{equation}
  \mu(h)= \exp\left\{
  \int_{h_0}^h\left[2+\Upsilon(h')\right]\,dh'\right\}.
  \label{e:mu_h_integral}
\end{equation}
Equations~(\ref{e:p_h_integral})--(\ref{e:mu_h_integral}) determine a
causal enthalpy-based equation of state generated by any non-negative
velocity function $\Upsilon(h)\geq 0$.

In this study two different types of parametric representations are
used.  The first type, piecewise-analytic representations, break the
relevant domain of enthalpies into $N_\mathrm{parms}$ subdomains, and
then expresses $\epsilon(h,\upsilon_a)$ and $p(h,\upsilon_a)$ in each
subdomain as a particular analytic function determined by the
parameters.  The particular causal piecewise analytic representations
used here are described in
Sec.~\ref{s:PiecewiseAnalyticRepresentations}.  The second type of
parametric representation used in this study is a spectral
representation that constructs $\epsilon(h,\upsilon_a)$ and
$p(h,\upsilon_a)$ from a generating function determined by a linear
combination of spectral basis functions.  The particular causal
spectral representation used here is described in
Sec.~\ref{s:CausalSpectralRepresentations}.

\subsection{ Causal Piecewise-Analytic Representations}
\label{s:PiecewiseAnalyticRepresentations}

The first step in constructing the causal piecewise-analytic
enthalpy-based representations used in this study is to divide the
enthalpy domain relevant for the high density portion of a
neutron-star core, $[h_\mathrm{min},h_\mathrm{max}]$, into
$N_\mathrm{parms}$ subdomains with $h_\mathrm{min}=h_0 < h_1 < ... <
h_{n-1} < h_{N_\mathrm{parms}}=h_\mathrm{max}$.  The representation
used in this study makes the subdomains uniformly spaced in $\log h$:
$\log(h_{a+1}/h_a)=N_\mathrm{parms}^{-1}\log(h_\mathrm{min}/h_\mathrm{max})$
for all $1\leq a \leq N_\mathrm{parms}$.

The second step is to choose analytical functions
$\Upsilon(h,\upsilon_a)$ to approximate $\Upsilon(h)$ in each
subdomain.  The challenge is to find analytical functions that are
reasonably good approximations in each subdomain, and that are simple
enough to allow Eqs.~(\ref{e:p_h_integral}) and
(\ref{e:epsilon_h_integral}) to be solved analytically for
$\epsilon(h,\upsilon_a)$ and $p(h,\upsilon_a)$.  Graphs in
Ref.~\cite{Lindblom2018} show that $\log\Upsilon$ is more or less
proportional to $\log h$ for a collection of model neutron-star
equations of state.  This fact, together with the need to have simple
functions that can be integrated analytically, lead to the following
choice for $\Upsilon(h,\upsilon_a)$~\cite{Lindblom2018},
\begin{eqnarray}
  \Upsilon(h,\upsilon_a)=\frac{\upsilon_{a}+2(h_{a}-h)}{h},
  \label{e:UpsilonhDef}
\end{eqnarray}
in the subdomain $h_{a-1}\leq h < h_{a}$.  These velocity functions are
non-negative within each subdomain so long as the adjustable
parameters are chosen to be non-negative, $\upsilon_a\geq 0$.

The piecewise-analytic representation of the equation of state,
$\epsilon(h,\upsilon_a)$ and $p(h,\upsilon_a)$, that corresponds to
the $\Upsilon(h,\upsilon_a)$ given in Eq.~(\ref{e:UpsilonhDef}) is
determined by evaluating the integrals in
Eqs.~(\ref{e:p_h_integral})--(\ref{e:mu_h_integral}). Inserting the
expression for $\Upsilon(h,\upsilon_a)$ from Eq.~(\ref{e:UpsilonhDef})
into these integrals gives the following expressions for the equation
of state,
\begin{eqnarray}
  &&\!\!\!\!\!\!
  p(h,\upsilon_a)=
  p_a+\frac{\left(\epsilon_a\,c^2+p_a\right)h_a}
  {\lambda_a+1}
  \left[\left(\frac{h}{h_a}\right)^{\lambda_a+1}-1\right],\nonumber\\
  &&\\
  &&\!\!\!\!\!\!
  \epsilon(h,\upsilon_a)=
  -p(h,\upsilon_a)c^{-2}+\left(\epsilon_a+p_a c^{-2}\right)
  \left(\frac{h}{h_a}\right)^{\lambda_a}\!\!\!\!,\quad
\end{eqnarray}
in the subdomain $h_a\leq h < h_{a+1}$, where
\begin{equation}
  \lambda_a=\upsilon_{a+1}+2h_{a+1}.
  \label{e:lambdaDef}
\end{equation}
The constants $p_a=p(h_a,\upsilon_a)$ and
$\epsilon_a=\epsilon(h_a,\upsilon_a)$ are determined from the
recursion relations,
\begin{eqnarray}
  &&p_{a+1}=
  p_a+\frac{\left(\epsilon_a\,c^2+p_a\right)h_a}{\lambda_a+1}
  \left[\left(\frac{h_{a+1}}{h_a}\right)^{\lambda_a+1}\!\!\!\!
    -1\right],\quad\nonumber\\
  &&\\
  &&\epsilon_{a+1}=
  -p_{a+1}c^{-2}+\left(\epsilon_a+p_a c^{-2}\right)
  \left(\frac{h_{a+1}}{h_a}\right)^{\lambda_a}.
\end{eqnarray}
The constants $p_0=p(h_0)\geq 0$ and $\epsilon_0=\epsilon(h_0)\geq 0$
are determined from the low-density equation of state at the matching
point $h=h_0$..

\subsection{Causal Spectral Representations}
\label{s:CausalSpectralRepresentations}

Spectral methods are very efficient ways to represent smooth functions,
providing good accuracy with only a small number of spectral basis
functions. This study is designed to test how well spectral
representations are able to represent non-smooth functions,
i.e. equations of state with phase transitions.  Causal spectral
representations are generated by a spectral expansions of the velocity
function $\Upsilon(h)$:
\begin{equation}
  \Upsilon(h)=\exp\left\{\sum_{a=1}^{N_\mathrm{parms}}\upsilon_a \Phi_a(h)\right\},
  \label{e:SpectralExpansion}
\end{equation}
where $\Phi_a(h)$ are a suitable set of spectral basis functions and
the constants $\upsilon_a$ are the spectral coefficients.  Inserting
this expression for $\Upsilon(h)$ into
Eqs.~(\ref{e:p_h_integral})--(\ref{e:mu_h_integral}) produces a causal
equation of state determined by the parameters $\upsilon_a$. Any
equation of state constructed in this way automatically produces a
velocity function $\Upsilon(h)$ that satisfies the causality condition
$\Upsilon(h)\geq 0$.

This study uses the very simple choice of spectral basis functions
$\Phi_a(h)=\left[\log(h/h_0)\right]^a$, which creates a collection of
velocity functions, $\Upsilon(h,\upsilon_a)$, parameterized by
$\upsilon_a$~\cite{Lindblom2022}:
\begin{equation}
  \Upsilon(h,\upsilon_a)=\Upsilon_0\exp\left\{\,
  \sum_{a=1}^{N_\mathrm{parms}}\,\upsilon_a
  \left[\log\left(\frac{h}{h_0}\right)\right]^a\right\}.
  \label{e:Upsilon_spectral}
\end{equation}
The constant $\Upsilon_0=\Upsilon(h_0)$ in this expression is
evaluated from the low-density portion of the equation of state at the
point $h_0$ using Eq.~(\ref{e:Upsilon_i}). Every choice of spectral
parameters $\upsilon_a$ in Eq.~(\ref{e:Upsilon_spectral}) determines a
non-negative velocity function, and using Eqs.~(\ref{e:p_h_integral})
and (\ref{e:epsilon_h_integral}) this generating function determines a
parameterized enthalpy-based causal equation of state,
$\epsilon=\epsilon(h,\upsilon_a)$ and $p=p(h,\upsilon_a)$.  These
integrals can not be done analytically, however, the integrands are
smooth and they can be evaluated numerically very accurately and
efficiently using Gaussian quadrature.  If the spectral expansion in
Eq.~(\ref{e:SpectralExpansion}) is convergent, then every causal
equation of state can be represented in this way by including enough
terms in the spectral expansion, i.e. by choosing $N_\mathrm{parms}$
sufficiently large.  However as this study shows, the spectral
representations of equations of state with large phase transitions are
not convergent for representations based on the particular spectral
expansion in Eq.~(\ref{e:Upsilon_spectral}). 

\vspace{1.5cm}

\break

\acknowledgments

I thank Jimmy Zhou for helping me debug parts of the code used in this
study, and Steve Lewis for numerous conversations related to this
study, and for helpful comments and suggestions on an earlier draft of
this paper.  This research was supported in part by the National
Science Foundation grant 2012857 to the University of California at
San Diego.  \vfill

\bibstyle{prd}
\bibliography{../References/References}

\begin{thebibliography}{11}
\expandafter\ifx\csname natexlab\endcsname\relax\def\natexlab#1{#1}\fi
\expandafter\ifx\csname bibnamefont\endcsname\relax
  \def\bibnamefont#1{#1}\fi
\expandafter\ifx\csname bibfnamefont\endcsname\relax
  \def\bibfnamefont#1{#1}\fi
\expandafter\ifx\csname citenamefont\endcsname\relax
  \def\citenamefont#1{#1}\fi
\expandafter\ifx\csname url\endcsname\relax
  \def\url#1{\texttt{#1}}\fi
\expandafter\ifx\csname urlprefix\endcsname\relax\def\urlprefix{URL }\fi
\providecommand{\bibinfo}[2]{#2}
\providecommand{\eprint}[2][]{\url{#2}}

\bibitem[{\citenamefont{Lindblom}(1992)}]{Lindblom1992}
\bibinfo{author}{\bibfnamefont{L.}~\bibnamefont{Lindblom}},
  \bibinfo{journal}{Astrophys.\ J.} \textbf{\bibinfo{volume}{398}},
  \bibinfo{pages}{569} (\bibinfo{year}{1992}).

\bibitem[{\citenamefont{Read et~al.}(2009)\citenamefont{Read, Lackey, Owen, and
  Friedman}}]{Read:2008iy}
\bibinfo{author}{\bibfnamefont{J.~S.} \bibnamefont{Read}},
  \bibinfo{author}{\bibfnamefont{B.~D.} \bibnamefont{Lackey}},
  \bibinfo{author}{\bibfnamefont{B.~J.} \bibnamefont{Owen}}, \bibnamefont{and}
  \bibinfo{author}{\bibfnamefont{J.~L.} \bibnamefont{Friedman}},
  \bibinfo{journal}{Phys. Rev.} \textbf{\bibinfo{volume}{D79}},
  \bibinfo{pages}{124032} (\bibinfo{year}{2009}).

\bibitem[{\citenamefont{Lindblom}(2010)}]{Lindblom2010}
\bibinfo{author}{\bibfnamefont{L.}~\bibnamefont{Lindblom}},
  \bibinfo{journal}{Phys.\ Rev.\ D} \textbf{\bibinfo{volume}{82}},
  \bibinfo{pages}{103011} (\bibinfo{year}{2010}).

\bibitem[{\citenamefont{Lindblom}(2018)}]{Lindblom2018}
\bibinfo{author}{\bibfnamefont{L.}~\bibnamefont{Lindblom}},
  \bibinfo{journal}{Phys.\ Rev.\ D} \textbf{\bibinfo{volume}{97}},
  \bibinfo{pages}{123019} (\bibinfo{year}{2018}).

\bibitem[{\citenamefont{Lindblom}(2022)}]{Lindblom2022}
\bibinfo{author}{\bibfnamefont{L.}~\bibnamefont{Lindblom}},
  \bibinfo{journal}{Phys.\ Rev.\ D} \textbf{\bibinfo{volume}{105}},
  \bibinfo{pages}{063031} (\bibinfo{year}{2022}).

\bibitem[{\citenamefont{Typel et~al.}(2010)\citenamefont{Typel, Ropke, Klahn,
  Blaschke, and Wolter}}]{Typel2010}
\bibinfo{author}{\bibfnamefont{S.}~\bibnamefont{Typel}},
  \bibinfo{author}{\bibfnamefont{G.}~\bibnamefont{Ropke}},
  \bibinfo{author}{\bibfnamefont{T.}~\bibnamefont{Klahn}},
  \bibinfo{author}{\bibfnamefont{D.}~\bibnamefont{Blaschke}}, \bibnamefont{and}
  \bibinfo{author}{\bibfnamefont{H.~H.} \bibnamefont{Wolter}},
  \bibinfo{journal}{Physical Review C} \textbf{\bibinfo{volume}{81}},
  \bibinfo{pages}{015803} (\bibinfo{year}{2010}).

\bibitem[{\citenamefont{Oppenheimer and Volkoff}(1939)}]{Oppenheimer1939}
\bibinfo{author}{\bibfnamefont{J.~R.} \bibnamefont{Oppenheimer}}
  \bibnamefont{and} \bibinfo{author}{\bibfnamefont{G.~M.}
  \bibnamefont{Volkoff}}, \bibinfo{journal}{Phys. Rev.}
  \textbf{\bibinfo{volume}{55}}, \bibinfo{pages}{374} (\bibinfo{year}{1939}).

\bibitem[{\citenamefont{Press et~al.}(1992)\citenamefont{Press, Teukolsky,
  Vetterling, and Flannery}}]{numrec_f}
\bibinfo{author}{\bibfnamefont{W.~H.} \bibnamefont{Press}},
  \bibinfo{author}{\bibfnamefont{S.~A.} \bibnamefont{Teukolsky}},
  \bibinfo{author}{\bibfnamefont{W.~T.} \bibnamefont{Vetterling}},
  \bibnamefont{and} \bibinfo{author}{\bibfnamefont{B.~P.}
  \bibnamefont{Flannery}}, \emph{\bibinfo{title}{Numerical Recipes in
  {FORTRAN}}} (\bibinfo{publisher}{Cambridge University Press},
  \bibinfo{address}{Cambridge, England}, \bibinfo{year}{1992}),
  \bibinfo{edition}{2nd} ed.

\bibitem[{\citenamefont{{Baym} et~al.}(1971)\citenamefont{{Baym}, {Pethick},
  and {Sutherland}}}]{BPS1971}
\bibinfo{author}{\bibfnamefont{G.}~\bibnamefont{{Baym}}},
  \bibinfo{author}{\bibfnamefont{C.}~\bibnamefont{{Pethick}}},
  \bibnamefont{and}
  \bibinfo{author}{\bibfnamefont{P.}~\bibnamefont{{Sutherland}}},
  \bibinfo{journal}{\apj} \textbf{\bibinfo{volume}{170}}, \bibinfo{pages}{299}
  (\bibinfo{year}{1971}).

\bibitem[{\citenamefont{Lindblom and Indik}(2012)}]{Lindblom2012}
\bibinfo{author}{\bibfnamefont{L.}~\bibnamefont{Lindblom}} \bibnamefont{and}
  \bibinfo{author}{\bibfnamefont{N.~M.} \bibnamefont{Indik}},
  \bibinfo{journal}{Phys.\ Rev.\ D} \textbf{\bibinfo{volume}{86}},
  \bibinfo{pages}{084003} (\bibinfo{year}{2012}).

\bibitem[{\citenamefont{Lindblom}(1998)}]{Lindblom98a}
\bibinfo{author}{\bibfnamefont{L.}~\bibnamefont{Lindblom}},
  \bibinfo{journal}{Phys.\ Rev.\ D} \textbf{\bibinfo{volume}{58}},
  \bibinfo{pages}{024008} (\bibinfo{year}{1998}).

\end{thebibliography}

\end{document}